\documentclass[trackchanges,preprint2]{aastex62}

\usepackage{amsmath}
\usepackage{breqn}
\usepackage{verbatim}

\newcommand{\tk}{t_{\rm kick}}
\newcommand{\Mbh}{M_{\mathrm{BH}}}
\newcommand{\vbh}{v_{\mathrm{BH}}}

\graphicspath{{./}{Figures/}}

\submitjournal{ApJL}

\shorttitle{The Beginning of an END}
\shortauthors{Akiba \& Madigan}

\begin{document}

\title{On the Formation of an Eccentric Nuclear Disk following the Gravitational Recoil Kick of a Supermassive Black Hole}

\author{Tatsuya Akiba}
\affiliation{JILA and Department of Astrophysical and Planetary Sciences, CU Boulder, Boulder, CO 80309, USA}
\email{tatsuya.akiba@colorado.edu}

\author{Ann-Marie Madigan}
\affiliation{JILA and Department of Astrophysical and Planetary Sciences, CU Boulder, Boulder, CO 80309, USA}

\begin{abstract}

The anisotropic emission of gravitational waves during the merger of two supermassive black holes can result in a recoil kick of the merged remnant. 
We show here that eccentric nuclear disks \--- stellar disks of eccentric, apse-aligned  orbits \--- can directly form as a result. An initially circular disk of stars will  align orthogonal to the black hole kick direction with a distinctive `tick-mark' eccentricity distribution and a spiral pattern in mean anomaly. 

\end{abstract}

\keywords{Galaxy nuclei \-- Astrodynamics \-- $N$-body problem \--  Gravitational waves}

\section{Introduction} \label{sec:intro}

The double nucleus of the Andromeda galaxy is well-explained by a massive disk of apse-aligned stellar orbits about a supermassive black hole \citep{Tremaine1995}, a configuration we call an eccentric nuclear disk (END). \citet{Madigan2018} showed that ENDs are stabilized by inter-orbit, gravitational torques that limit differential apsidal precession and \citet{Gruzinov2020} demonstrated that they are a thermodynamic equilibrium configuration for rotating nuclear stellar clusters. These eccentric, lopsided stellar disks may turn out to be fairly common in galactic nuclei.

Tidal disruption events (TDEs) occur when stars are unbound by a black hole's tidal gravity \citep{Hil75}. {TDEs are observed in post-starburst galaxies at higher rates than other galaxy types \citep{Arcavi2014, Stone2016b, French2016} and these starbursts are likely linked to galaxy mergers \citep{Barnes91, Cales2015,Hammerstein2021}.} Stellar populations in ENDs can produce TDE rates $10^3 - 10^4$ times that of a spherical, isotropic configuration \citep{Madigan2018}. We propose a formation mechanism for ENDs which links galaxy mergers to enhanced TDE rates.

Binary supermassive black holes can form as a result of galaxy mergers \citep{Beg80}. Gravitational waves are emitted during their inspiral and eventual coalescence \citep{Einstein1916}. In an unequal-mass binary system, the less massive black hole has a higher orbital velocity and is thus more effective at forward-beaming its gravitational wave radiation \citep{Wiseman92}. This results in anisotropic gravitational waves which carry away linear momentum and the remnant receives a recoil kick \citep{Bekenstein1973}. The same effect is produced for spinning black hole binaries \citep{Herrmann07}. In rare cases, the kick velocity may exceed the escape velocity of the galaxy leading to black hole ejection \citep{Gualandris2008}, but even for smaller kick velocities, there are important dynamical consequences \citep{Blecha}. Here we show that ENDs can form as a result of a recoil kick, explaining the enhanced TDE rates we observe in post-merger galaxies.

In Section \ref{sec:analytics}, we introduce a toy model to analytically explain apsidal alignment that may result from a black hole recoil kick. In Section \ref{sec:n-body}, we confirm this formation mechanism using $N$-body simulations and discuss the orbital properties of the resultant ENDs. We summarize our findings and conclude in Section \ref{sec:conclusion}.

\section{Formation Mechanism} \label{sec:analytics}

\begin{figure*}
\centering
\begin{tabular}{cc}
\includegraphics[width=0.3\linewidth]{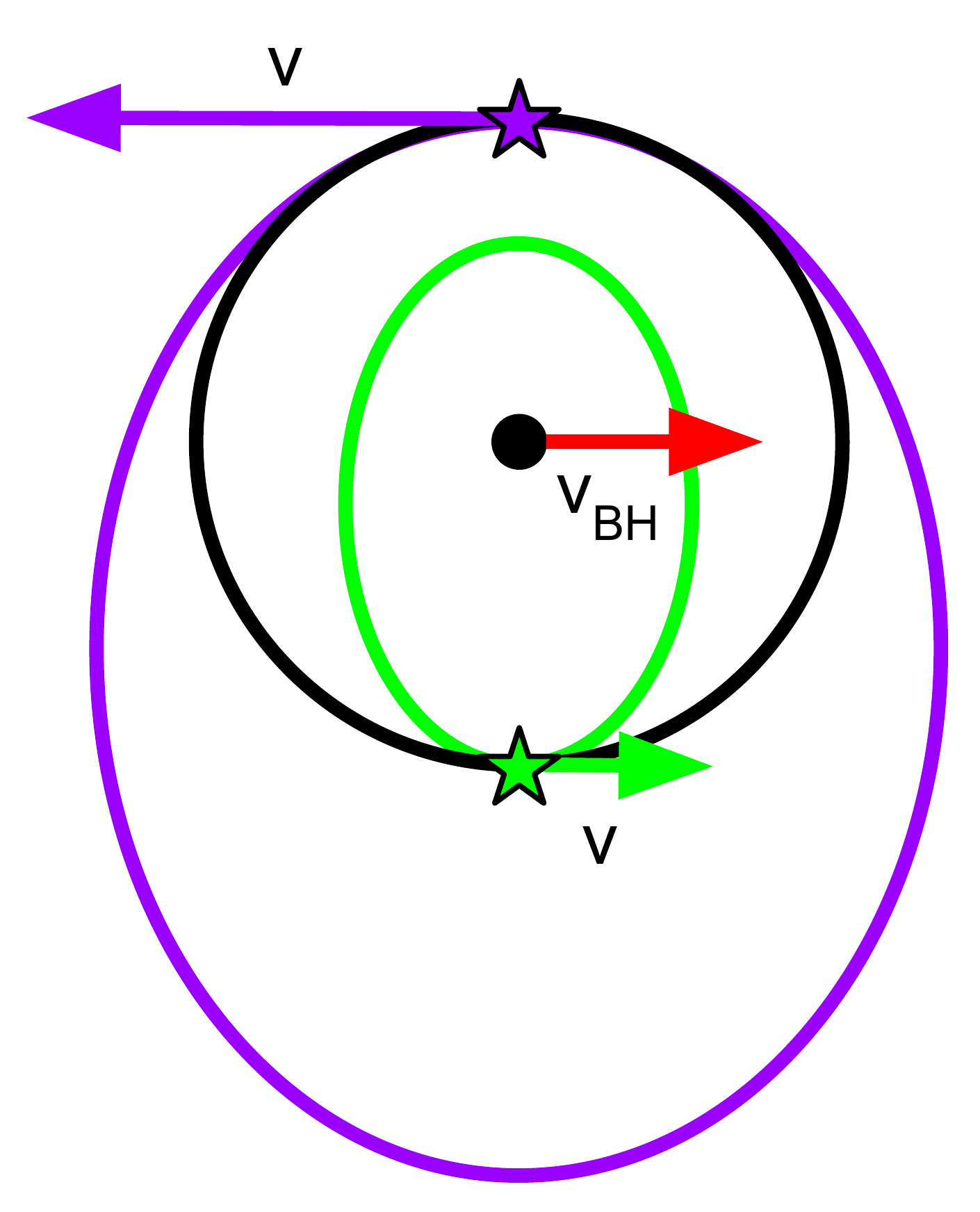} &
\includegraphics[width=0.49\linewidth]{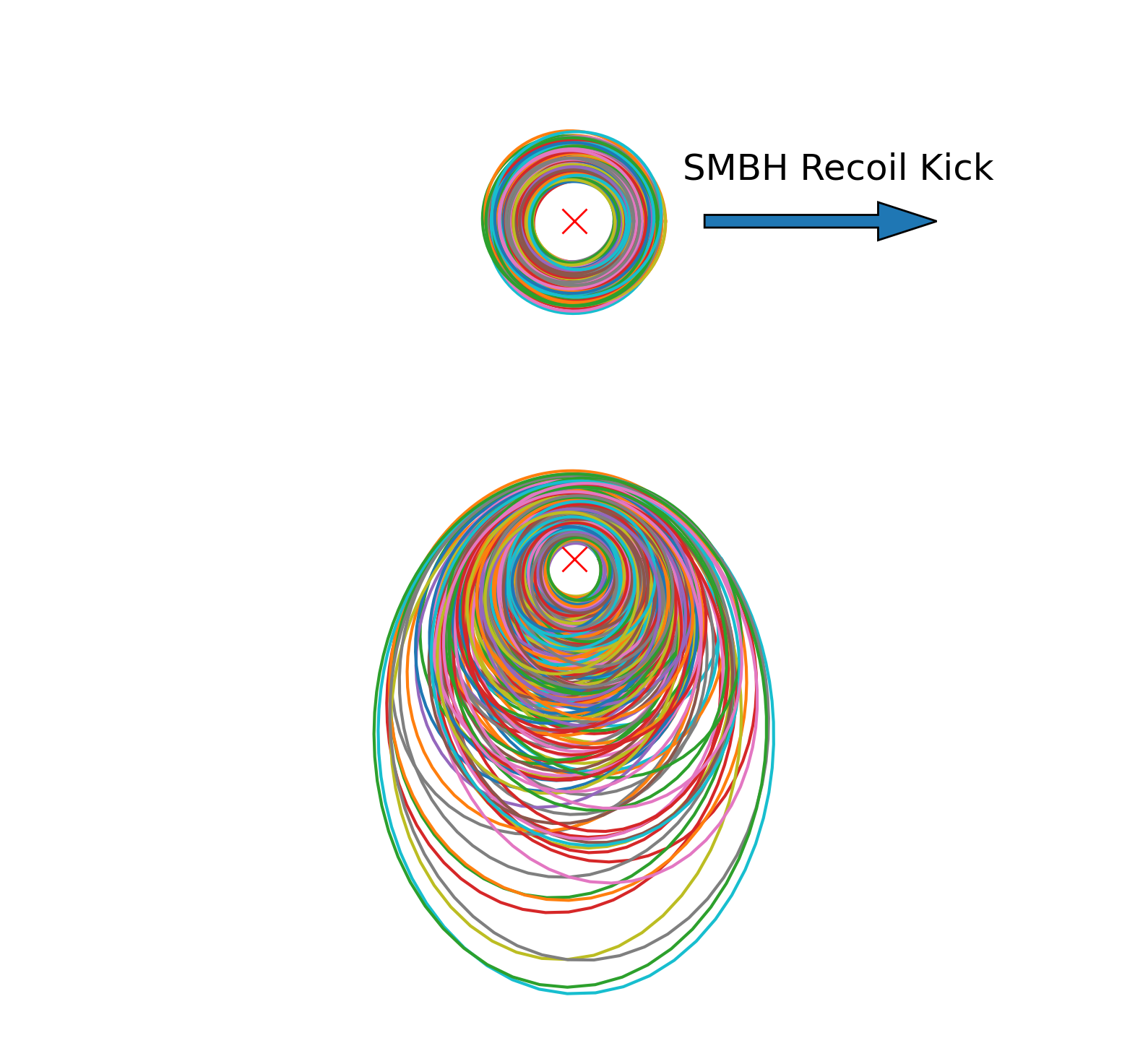} \\
(a) Schematic diagram. & (b) $N$-body result. \\
\end{tabular}
    \caption{END formation through a black hole recoil kick. In the schematic (a), two stars initially move along a circular orbit shown by the black line. Post-kick, the top (purple) star lies on a more eccentric, larger orbit and the bottom (green) star lies on an eccentric, smaller orbit. Both eccentricity vectors are aligned orthogonal to the black hole kick. In the $N$-body result (b), the stellar orbits before (top) and immediately following (bottom) the black hole kick are shown. The position of the black hole is marked with a red cross and the direction of kick marked with a blue arrow.}
    \label{fig:formation}
\end{figure*}

Here we set up a toy problem that allows us to predict the orbital structure of a disk of stars following a kick to the central black hole. 
We consider the simple case of a circular disk formed from a self-gravitating accretion disk \citep{Pac78,Goo03} and focus on an in-plane recoil kick motivated by \citet{Bogdanovic2007} (but see also \citet{Lodato2013}).
To measure apsidal alignment we use the eccentricity vector, 
\begin{equation}
    \vec{e} = \frac{\vec{v} \times \vec{j}}{G\Mbh} - \frac{\vec{r}}{r} \ ,
    \label{eqn:eccentricity_vector}
\end{equation}
\noindent where $\vec{r}$ is the radius vector, $\vec{v}$ is the velocity vector, $\vec{j}$ is the angular momentum vector, and $\Mbh$ is the mass of the central black hole. The eccentricity vector points from the apoapsis to the periapsis of an orbit and its magnitude is equal to the scalar eccentricity. 

We consider a circular orbit in the  $x$-$y$ plane with positive angular momentum ($|j| = j_z$) centered on the black hole located at the origin. Pre-kick, the star's initial position and velocity are given by

\begin{equation}
\vec{r} = r \cos(\Omega t) \ \hat{x} + r \sin(\Omega t) \ \hat{y} \ ,
\label{eqn:position}
\end{equation}

\begin{equation}
\vec{v} = - v_{\mathrm{circ}} \sin(\Omega t) \ \hat{x} + v_{\mathrm{circ}} \cos(\Omega t) \ \hat{y} \ ,
\label{eqn:velocity}
\end{equation}

\noindent where $\Omega  \equiv (\frac{G\Mbh}{r^3})^{1/2}$ is the angular velocity, and $v_{\mathrm{circ}} \equiv \Omega r$ is the circular velocity at this radius. Without loss of generality, we give the black hole a kick in the positive $x$-direction with magnitude $\vbh$ at $t = \tk$. We define the initial anomaly of the star at this instant as $\theta \equiv \Omega \tk$. Immediately following the kick, the position of the star remains unchanged but in the reference frame of the black hole, its velocity becomes

\begin{equation}
\vec{v} = (- v_{\mathrm{circ}} \sin(\theta) - \vbh) \ \hat{x} + v_{\mathrm{circ}} \cos(\theta) \ \hat{y}.
\label{eqn:velocity_after}
\end{equation}

\noindent Setting the star's pre-kick eccentricity vector to zero, and using trigonometric identities,  the post-kick eccentricity vector becomes

\begin{dmath}
    \vec{e} = \frac{1}{G\Mbh} \left(\frac{1}{2} r v_{\mathrm{circ}} \vbh \sin(2 \theta) \ \hat{x} \ + 
    (r \vbh^2 \sin(\theta) + r v_{\mathrm{circ}} \vbh \sin^2(\theta) + r v_{\mathrm{circ}} \vbh) \ \hat{y} \right).
    \label{eqn:eccentricity_final_full}
\end{dmath}

\noindent To approximate a population of stars in an initially circular disk, we assume that the initial anomalies are evenly distributed in $[0, 2 \pi)$ radians. The average eccentricity vector then becomes

\begin{dmath}
 \langle \vec{e} \rangle = \frac{1}{2 \pi} \int_0^{2 \pi} \vec{e} (\theta) \ d \theta
 = \frac{3}{2} \frac{\vbh}{v_{\mathrm{circ}}} \ \hat{y}.
\label{eqn:average_eccentricity}
\end{dmath}

Figure \ref{fig:formation}(a) shows a schematic diagram representing two stars in this toy problem. We show the initial circular orbit in black with the kick direction indicated by the red arrow labeled $\vbh$. At the time of the kick, the top (purple) star's instantaneous circular velocity is pointed in the negative $x$-direction. In the reference frame of the black hole the star takes on a larger velocity as indicated by the top (purple) arrow, and moves onto a larger, eccentric orbit shown by the large (purple) ellipse. The bottom (green) star's instantaneous circular velocity is in the positive $x$-direction in the same direction as the kick. Thus, in the frame of the black hole the star slows down, takes on a smaller velocity as shown by the bottom (green) arrow, and moves onto a smaller, eccentric orbit indicated by the small (green) ellipse. Both ellipses have eccentricity vectors that point toward the positive $y$-direction, in agreement with Equation \ref{eqn:average_eccentricity}. This is analogous to the Hohmann transfer orbit, an orbital maneuver that is often used to transfer a spacecraft between two circular orbits of different radii \citep{Palmore84}. Black hole kicks induce apsidal alignment in nearby circular stellar orbits and, perhaps counter-intuitively, the eccentric disk forms in such a way that the net eccentricity vector direction is orthogonal to the kick direction. We confirm our analytic approach with $N$-body simulations in the next section.

\section{$N$-body Simulations} \label{sec:n-body}

\subsection{Numerical Set-Up} \label{sec:sim_setup}

We use the open-source, $N$-body simulation package \texttt{REBOUND} with the \texttt{IAS15} integrator, a high accuracy non-symplectic integrator with adaptive time-stepping \citep{Rein2012,Rein2015}. We use code units of $G = 1$, $\Mbh = 1$, and the inner edge of the disk $a_{\mathrm{in}} = 1$ such that the period of a circular orbit at the inner edge of the disk is $P(a_{\mathrm{in}}) = 2 \pi$.

We initialize stars in an axisymmetric, thin disk with $N = 400$ stars in the range $a = 1 - 2$ with surface density profile $\Sigma \propto a^{-2}$, eccentricities uniformly distributed in $e = 0 - 0.1$, inclination Rayleigh distributed with scale parameter $\sigma = 3^\circ$, longitude of periapsis ($\varpi$) and mean anomaly uniformly distributed in $[0 - 2 \pi)$, and disk mass $M_{\mathrm{disk}} = 10^{-2} \Mbh$.
At $\tk = 10$, we kick the black hole in the plane of the disk with varying magnitudes of $\vbh = (0.1 - 1.0) \ v_{\mathrm{circ}}(a_{\mathrm{out}})$ in increments of $0.1 \ v_{\mathrm{circ}}(a_{\mathrm{out}})$, where $v_{\mathrm{circ}}(a_{\mathrm{out}})$ is the circular velocity of stars at the outer edge of the disk ($a_{\mathrm{out}} = 2$). Scaling our code units loosely to the M31 system \citep{Lauer1993, Bender2005} with $\Mbh = 10^8 M_{\odot}$ and $a_{\mathrm{in}} = 1 \ \mathrm{pc}$ means $t = 1$ translates to $\sim 1490 \ \mathrm{yr}$. The simulation in Figure~\ref{fig:ave_unit_ecc_vec} then runs until $\sim 75$ Myr with a black hole kick of $\vbh = 0.4 \ v_{\mathrm{circ}}(a_{\mathrm{out}}) \approx 186 \ \rm{km/s}$.

\subsection{Formation of an END} \label{sec:results}

\indent In Figure \ref{fig:formation}(b), we show the stellar orbits projected into the $x$-$y$ plane before and after the recoil kick for a simulation with $\vbh = 0.3 \ v_{\mathrm{circ}}(a_{\mathrm{out}})$. The black hole is indicated by the red cross and the kick direction with a blue arrow. Immediately following the kick,  the orbits become more eccentric and the eccentricity vectors point in the positive $y$-direction as predicted.

\begin{figure}
\begin{center}
\includegraphics[width=\linewidth]{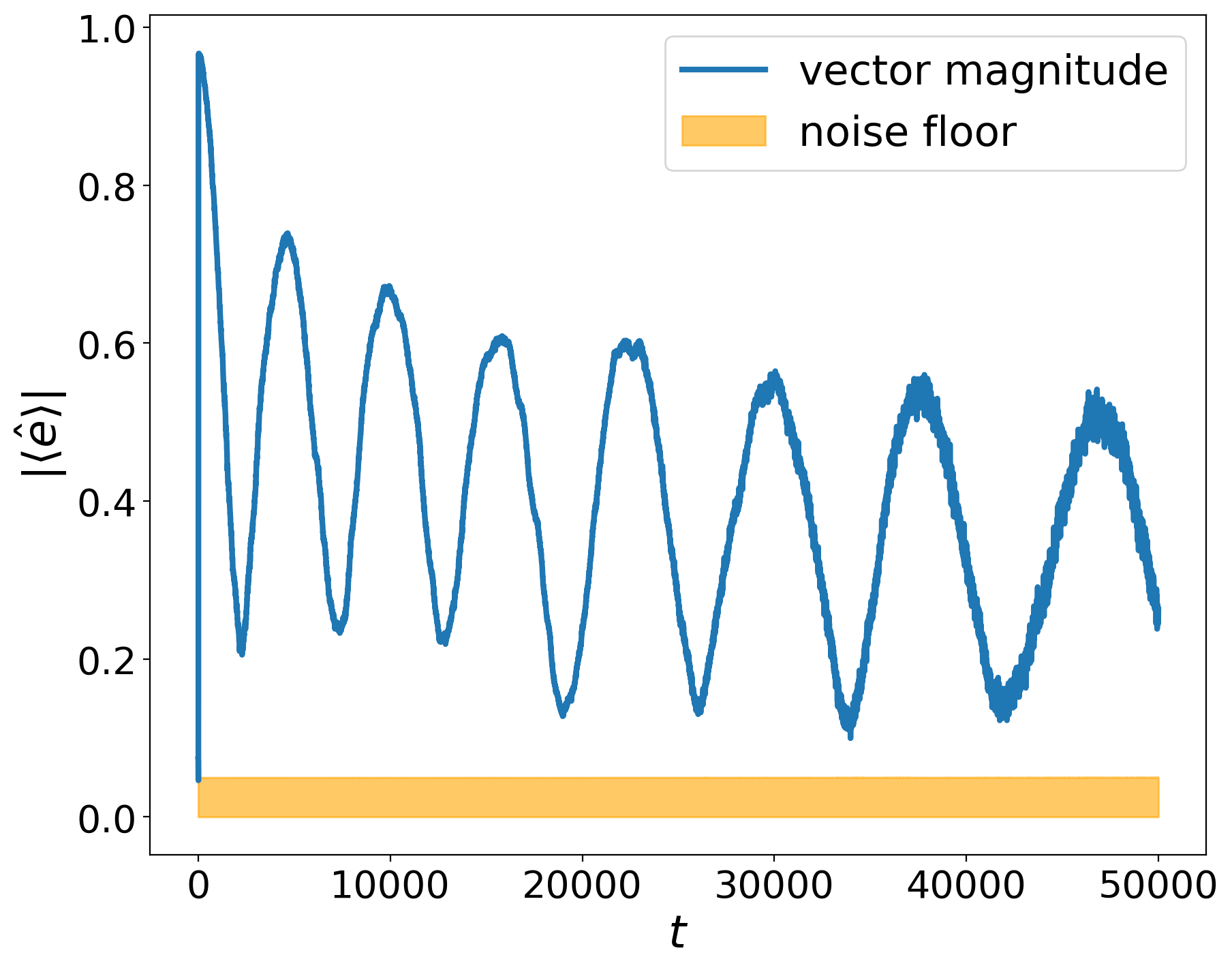}
\caption{Time evolution of the average unit eccentricity vector of stars in the disk during and after the black hole kick. The vector magnitude is shown with the solid (blue) line and the noise floor is shown in the (orange) shaded region.}
\label{fig:ave_unit_ecc_vec}
\end{center}
\end{figure}

We quantify the net apsidal alignment using the average unit eccentricity vector,

\begin{equation}
    \langle \hat{e} \rangle = \frac{ \sum_{i = 1}^{N_{\rm bound}} \hat{e}_i}{N_{\rm bound}},
\end{equation}

\noindent where $N_{\mathrm{bound}}$ is the total number of bound stellar particles, and $\hat{e}_i$ denotes the unit eccentricity vector of the $i$-th particle. In Figure \ref{fig:ave_unit_ecc_vec}, we show the time evolution of the magnitude of the average unit eccentricity vector of the disk stars for a simulation with $\vbh = 0.4 \ v_{\mathrm{circ}}(a_{\mathrm{out}})$. The black hole receives a kick at $\tk = 10$. The noise floor, calculated as $1/\sqrt{N_{\mathrm{bound}}}$, is shown in the (orange) shaded region. The average unit eccentricity vector lies below the noise floor before the kick. After the kick, it jumps to a magnitude of close to one indicating near perfect apsidal alignment. It then undergoes a coherent, large-amplitude oscillation. This behavior is well-explained by the alignment/mis-alignment of stars separated in semi-major axis space, as explained below.

\begin{figure*}
   \centering
\begin{tabular}{ccc}
\includegraphics[width=0.31\linewidth]{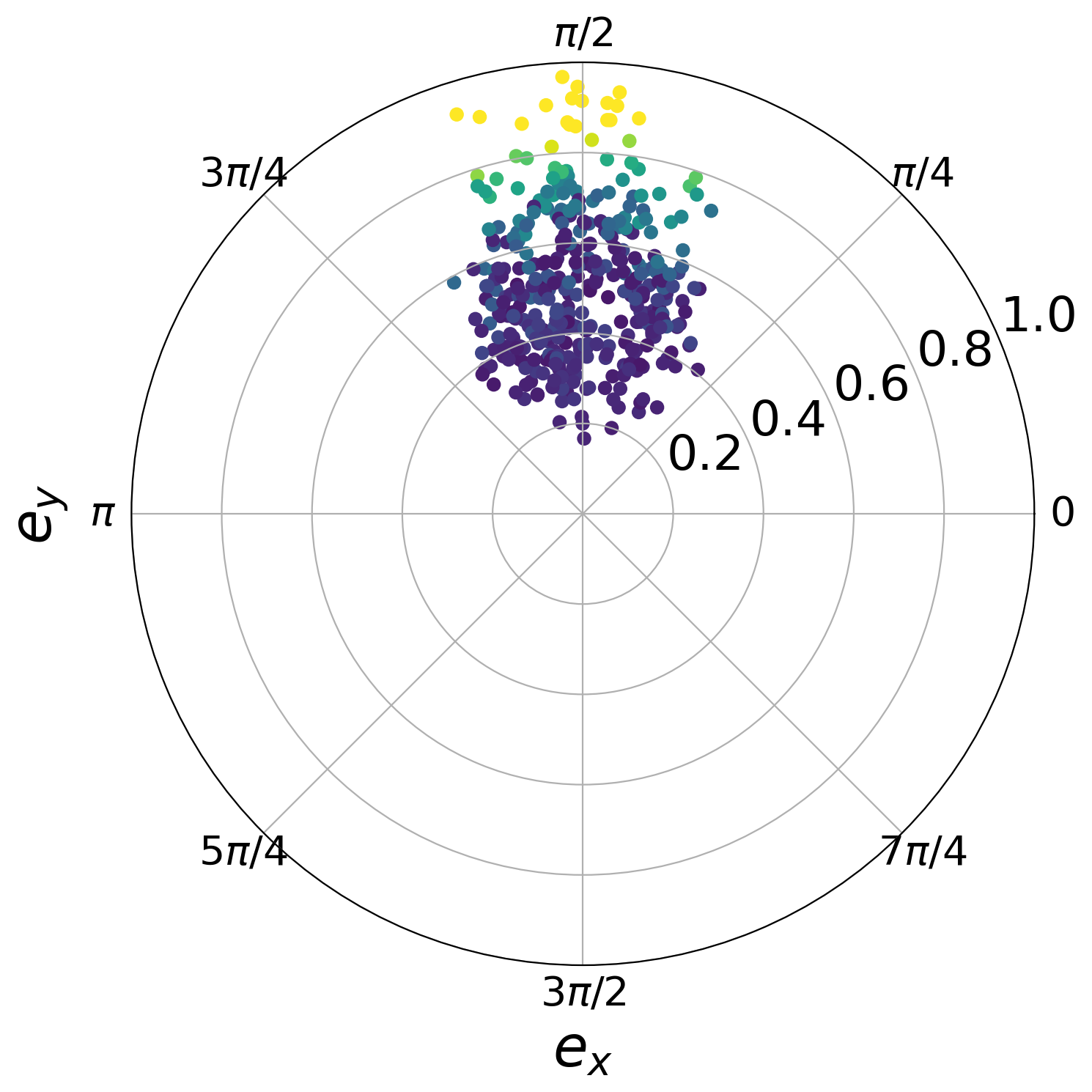} & \includegraphics[width=0.31\linewidth]{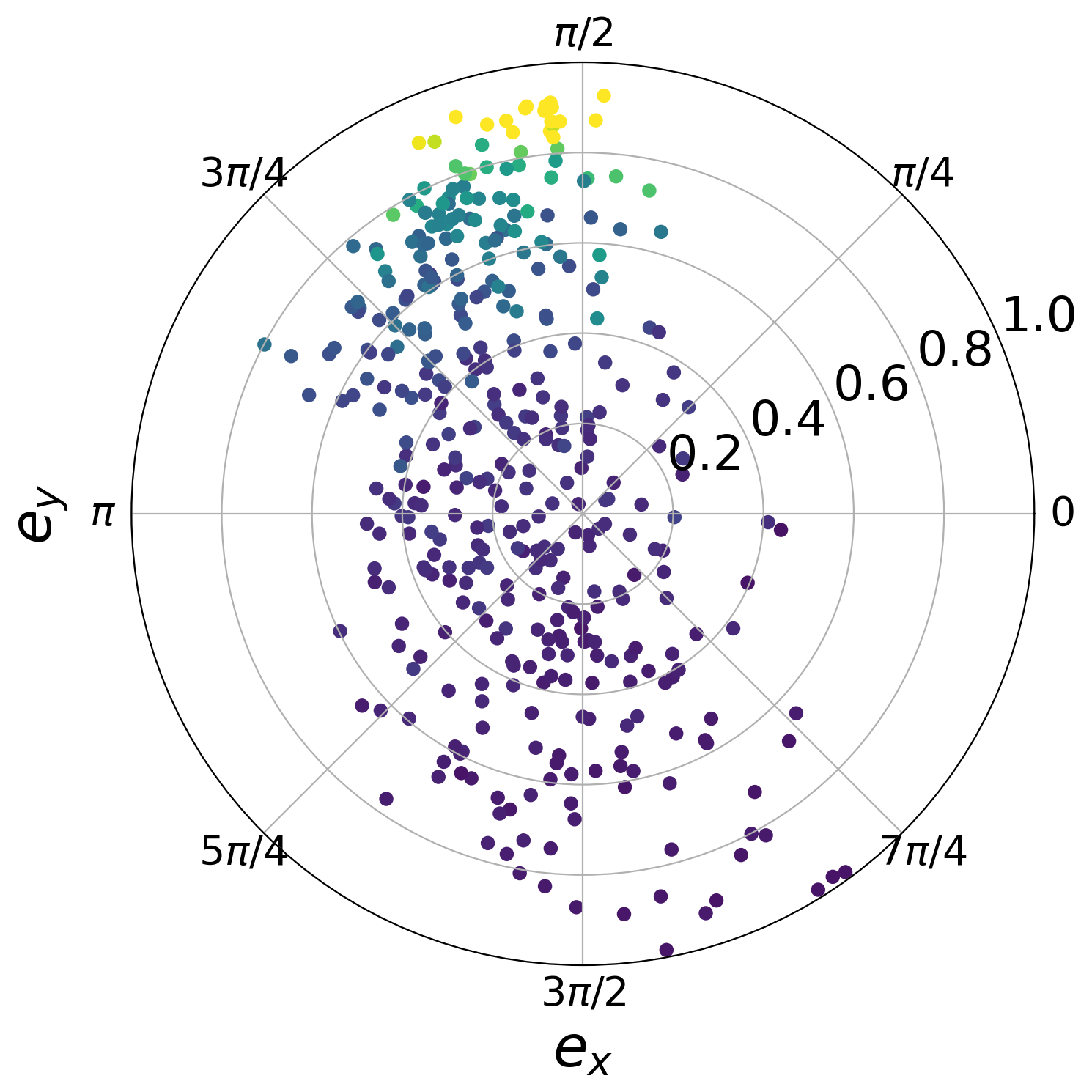} &
\includegraphics[width=0.36\linewidth]{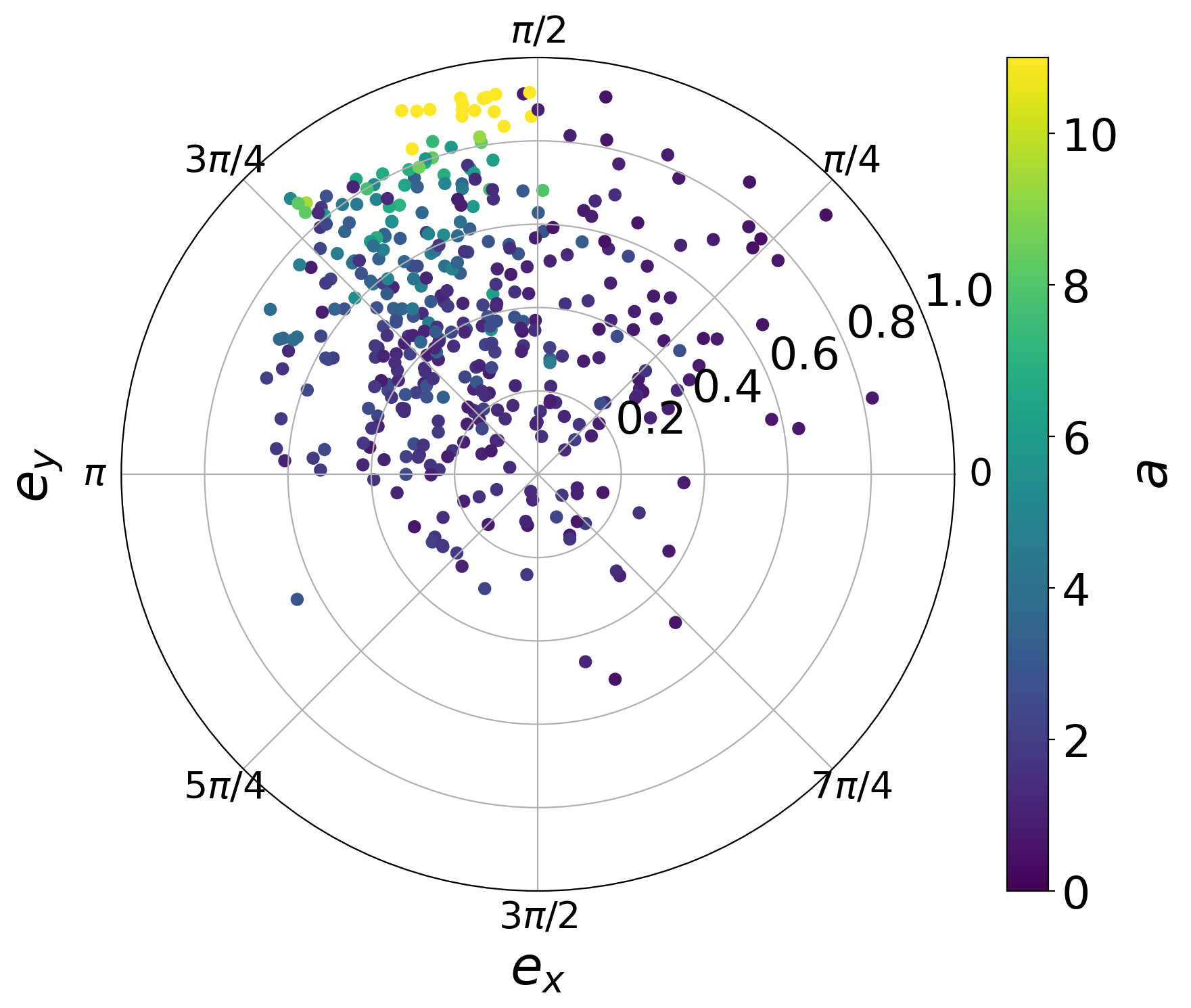} \\
(a) $t = \tk$. & (b) $t  = \tk + 1750$. & (c) $t = \tk + 4100$. \\
\end{tabular}
    \caption{Evolution of the in-plane eccentricity vectors $(e_x,e_y)$ of stars in an initially axisymmetric disk following a recoil kick of the central black hole. The color bar shows the semi-major axes of the stars.}
    \label{fig:ex-ey}
\end{figure*}

In Figure \ref{fig:ex-ey}, we show a polar plot of the eccentricity vector components ($e_x$, $e_y$) after the kick, with the semi-major axis indicated with the color bar. This figure explains the origin of the oscillation in apsidal alignment seen in Figure \ref{fig:ave_unit_ecc_vec}. 
Immediately following the kick (a), stellar orbits become eccentric and aligned in the positive $y$-direction. Stars with small semi-major axes precess much faster than those with larger semi-major axes, and misalignment (b) and re-alignment (c) between these groups occur periodically. 

\subsection{Eccentricity Gradient} \label{sec:ecc_grad}

\indent Specifying $\vbh$, $r$ and its corresponding $v_{\mathrm{circ}}$, the orbital eccentricity of a star post-kick depends only on the anomaly of the star at the time of the kick, $\theta$, which we see from taking the magnitude of the eccentricity vector in Equation~\ref{eqn:eccentricity_final_full}. The semi-major axis of the post-kick orbit can be determined from the vis-viva equation,

\begin{equation}
    v^2 = G\Mbh \left( \frac{2}{r} - \frac{1}{a} \right).
    \label{eqn:vis-viva}
\end{equation}

\noindent Substituting Equation~\ref{eqn:velocity_after} we obtain, 

\begin{multline}
    a (\theta) = {G\Mbh r}  \cdot [2G\Mbh -  \\
    ((-v_{\mathrm{circ}} \sin(\theta) - \vbh)^2 + (v_{\mathrm{circ}} \cos (\theta))^2)r]^{-1}.
    \label{eqn:sma_profile}
\end{multline}

\begin{figure*}
   \centering
\begin{tabular}{cc}
\includegraphics[width=0.45\linewidth]{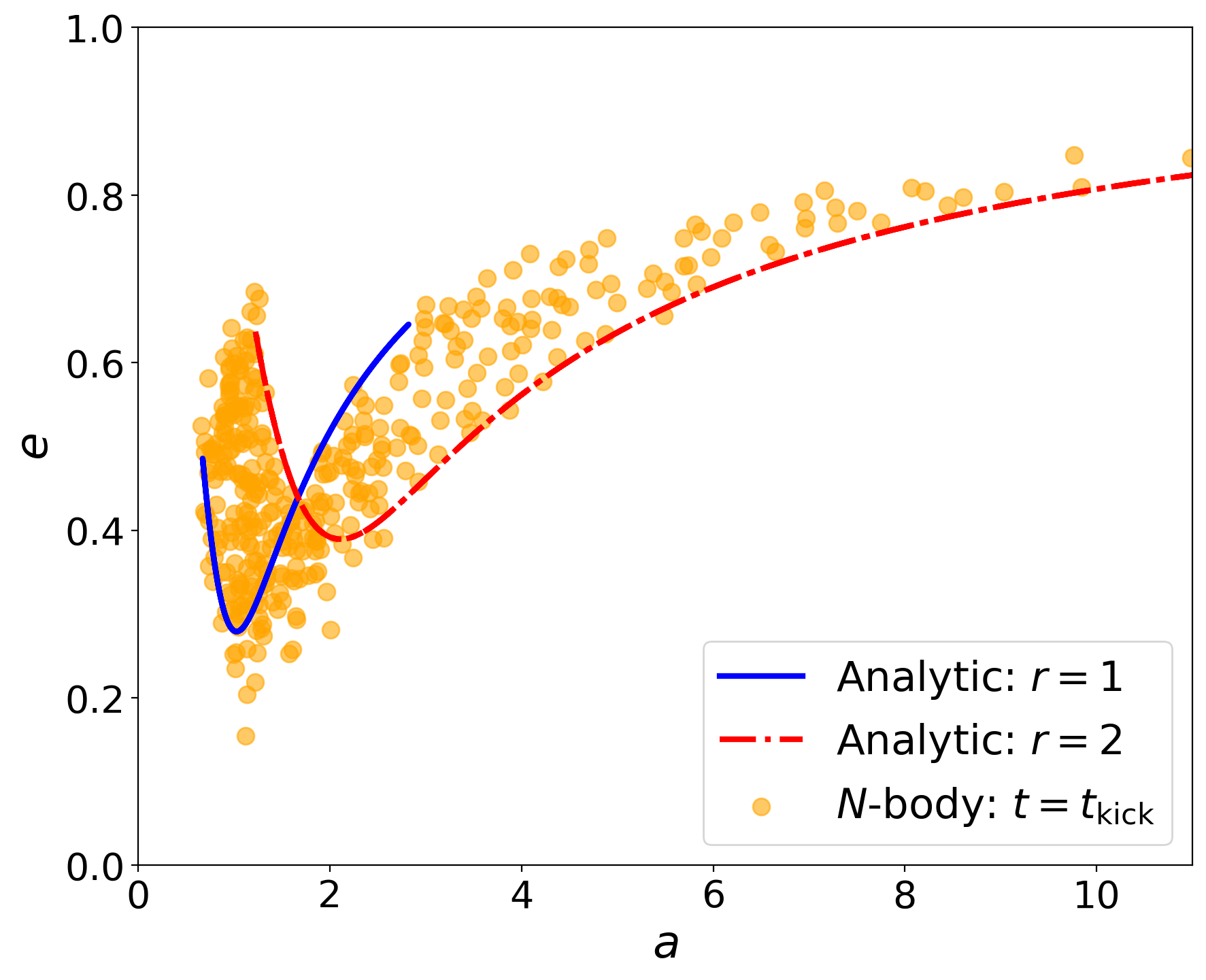} &
\includegraphics[width=0.45\linewidth]{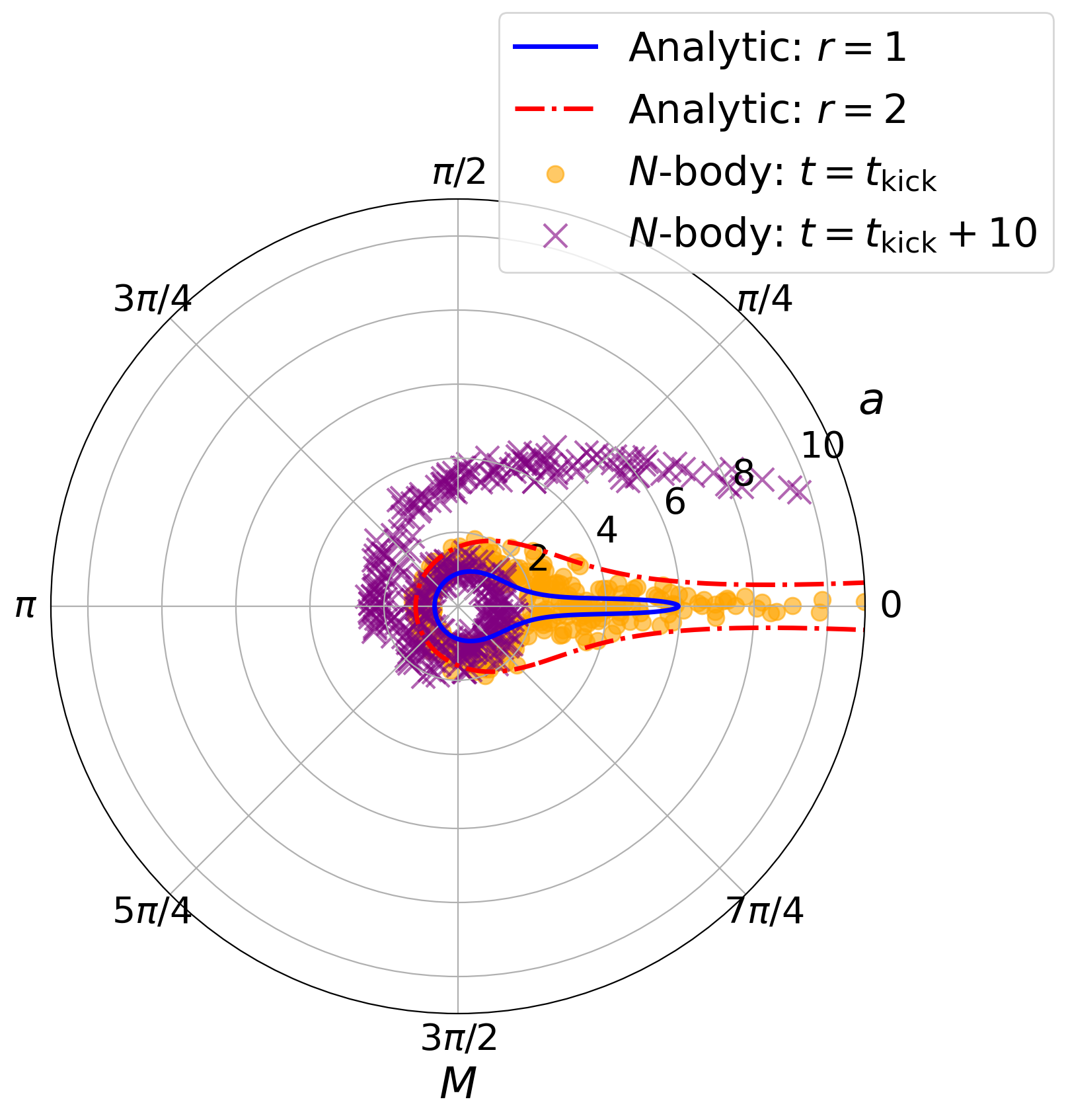} \\
(a) Eccentricity as a function of semi-major axis. & (b) Mean anomaly as a function of semi-major axis. \\
\end{tabular}
    \caption{Orbital structure of an eccentric nuclear disk following a black hole recoil kick. Analytic expectations for circular orbits at $r=1$ and $r=2$ are shown in solid (blue) and dash-dotted (red) lines. $N$-body results at $t = \tk$ are shown with orange circles, and at $t = \tk + 10$ with purple crosses.}
    \label{fig:orbital_properties}
\end{figure*}

\indent In Figure \ref{fig:orbital_properties}(a), we show the post-kick eccentricities as a function of semi-major axes in an $N$-body simulation with $\vbh = 0.4 \ v_{\mathrm{circ}}(a_{\mathrm{out}})$. In solid (blue) and dash-dotted (red) lines, we show the analytic expectation for circular orbits that begin at the inner edge ($r = 1$) and the outer edge ($r = 2$) of the disk, respectively. The  eccentricity profile has a distinctive `tick-mark' shape. There is a minimum at the radius corresponding to the initial semi-major axis of the stellar orbits and eccentricity values increase both inward or outward, such that the eccentricity gradient is negative for stars with smaller post-kick semi-major axes and positive for stars with larger post-kick semi-major axes. Whilst the orbits diffuse due to two-body relaxation, this pattern is maintained in our simulations for at least $t = 50,000$.

\subsection{Spiral in Mean Anomaly}

We observe a clear pattern in the distribution of mean anomalies of the stars in the disk following the kick. We can see why a pattern might emerge from Figure \ref{fig:formation}(a). The top (purple) star, which takes on a large semi-major axis, is positioned such that it is located at the periapsis of its new orbit. On the other hand, the bottom (green) star, which takes on a small semi-major axis, is positioned at the apoapsis of its new orbit. This suggests that there is a pattern in anomalies following the kick that is correlated with the stars' semi-major axes. We calculate true anomaly as,

\begin{equation}
    \nu = \begin{cases}
    \arccos \frac{\vec{e} \cdot \vec{r}}{|\vec{e}||\vec{r}|} , \ \ \ \ \ \ \ \ \ \ \mathrm{if} \ \vec{r} \cdot \vec{v} > 0 \\
    2 \pi - \arccos \frac{\vec{e} \cdot \vec{r}}{|\vec{e}||\vec{r}|} , \ \ \ \mathrm{if} \ \vec{r} \cdot \vec{v} < 0 \ .
    \end{cases}
    \label{eqn:true_anom}
\end{equation}

\noindent We then convert to mean anomaly, $M$, by a standard series expansion \citep{Smart1953}. Figure \ref{fig:orbital_properties}(b) shows results of an $N$-body simulation with $\vbh = 0.4 \ v_{\mathrm{circ}}(a_{\mathrm{out}})$ of mean anomalies as a function of semi-major axis immediately post-kick (orange circles) and shortly after at $t = \tk + 10$ (purple crosses). Once again, the expected distribution for initially circular orbits at $r=1$ and $r=2$ are shown by solid (blue) and dash-dotted (red) lines, respectively. A spiral structure emerges which continues to wind for several tens of orbital periods and dissipates in less than a precession period ($ \approx \Mbh/M_{\rm disk} \ P$).

\section{Discussion} \label{sec:conclusion}

In this letter we propose a new formation mechanism for eccentric nuclear disks (ENDs) via the gravitational wave recoil kick of the central black hole. We show that in-plane kicks cause surrounding circular stellar orbits to take on eccentricities orthogonal to the direction of the recoil kick. 
     Post-kick, a transient spiral structure emerges in mean anomaly as a function of semi-major axis. The orbits display a distinctive and long-lived `tick-mark' pattern in their eccentricity as a function of semi-major axis. 
On secular timescales we show that the average unit eccentricity vector undergoes a coherent oscillation due to the alignment/mis-alignment of stars separated in semi-major axis space (Figure~\ref{fig:ave_unit_ecc_vec}). 

In future work we will explore the stability of these aligned disks as a function of kick magnitude. Weak kicks will not result in significant apsidal alignment (Equation~\ref{eqn:average_eccentricity}), while extreme kicks will leave few stars bound to the black hole. 
A recoil kick with a significant in-plane component is required to generate an END from a circular disk of stars. In future work we will consider recoil kicks with out-of-plane components, as expected for spinning Kerr black holes \citep{Baker08}. 
We will explore other astrophysically-relevant stellar configurations including a scoured nuclear core \citep{Beg80} and consider the effects of gas in the system. 

\citet{StoneLoeb11, StoneLoeb12} show that recoil kicks temporarily enhance TDE rates up to $0.1 \ \mathrm{yr}^{-1} \ \mathrm{gal}^{-1}$ by instantaneously filling the central black hole's loss cone (see also \citet{Komossa2008}). Our preliminary results suggest that if an END forms,  elevated TDE rates are actually maintained for a long period of time (at least several precession periods). Such high rates could have a significant impact on the lifetime of ENDs formed via recoil kicks. However, much like bars in disk galaxies which are also composed of aligned stellar orbits, ENDs will evolve dynamically over time. They may dynamically attract new stellar orbits and grow in mass as they age. To this end, we will study the long-term evolution of ENDs formed in this manner. 

Finally, this work applies to many other astrophysical scenarios, for example, when the recoil kick is due to anisotropic mass loss from a star or its non-spherical collapse to a stellar mass black hole or neutron star (which leads us to contemplate an END of planets). 
In the neutron star case, typical kicks are on the order $\sim$380 km/s \citep{Faucher2006} which means that only the most tightly bound planets will be retained by the neutron star after an in-plane kick. 
White dwarf kicks on the other hand are of much smaller velocity. Kicks of order $\sim$1 km/s \citep{stone2015} may lead to END formation beyond the planetary regime at $\sim1000$ AU. In our Solar System, this relatively low mass region is populated by scattered disk objects on high eccentricity orbits. This low-kick-velocity, high orbital eccentricity configuration is one we will explore in future work.

\section{Acknowledgements} \label{sec:acknowledgements}

We thank Angela Collier for the short title suggestion and the anonymous referee for their suggestions which greatly improved the quality of our paper. AM gratefully acknowledges support from the David and Lucile Packard Foundation. This work utilized resources from the University of Colorado Boulder Research Computing Group, which is supported by the National Science Foundation (awards ACI-1532235 and ACI-1532236), the University of Colorado Boulder, and Colorado State University. \\

\software{\texttt{REBOUND} \citep{Rein2012}}

\bibliographystyle{aasjournal}
\bibliography{ms}

\end{document}